\documentclass[12pt]{article}
\pdfoutput=1
  \newcommand{\be}[1]{\begin{equation}\label{#1}}
  \newcommand{\ba}[1]{\begin{eqnarray}\label{#1}}
  \newcommand{\ep}[1]{\epsilon_{#1}}

  \newcommand{\rd}{{\rm d}}
  
  \newcommand{\re}{{\rm e}}
  \newcommand{\pa}[1]{\left(#1\right)}
  \newcommand{\paq}[1]{\left[#1\right]}
  \newcommand{\pag}[1]{\left\{#1\right\}}
  \newcommand{\av}[1]{\langle#1\rangle}
  \newcommand{\M}{{\rm M_{\rm P}}}
   \newcommand{\m}{{\rm m_{\rm P}}}

  \def\ee{\end{equation}}
  \def\ea{\end{eqnarray}}

\usepackage{amsmath}
\DeclareMathOperator\arctanh{arctanh}
\usepackage{graphicx,latexsym}
\usepackage{graphicx,epsf}
\usepackage{color}
\usepackage{epsfig}
\usepackage{amsmath}
\usepackage[T1]{fontenc}
\usepackage[utf8]{inputenc}
\usepackage{tikz}
\usepackage{pgfplots}
\usepackage{authblk}
\begin{document}
\title{The Born-Oppenheimer approach to Quantum Cosmology}
\author[1]{Alexander Y. Kamenshchik\thanks{Alexander.Kamenshchik@bo.infn.it}}
\author[2]{Alessandro Tronconi\thanks{Alessandro.Tronconi@bo.infn.it}}
\author[2]{Giovanni Venturi\thanks{Giovanni.Venturi@bo.infn.it}}
\affil[1]{Dipartimento di Fisica e Astronomia and INFN, Via Irnerio 46,40126 Bologna,
Italy\\

L.D. Landau Institute for Theoretical Physics of the Russian
Academy of Sciences, 119334 Moscow, Russia}
\affil[2]{Dipartimento di Fisica e Astronomia and INFN, Via Irnerio 46, 40126 Bologna,
Italy}

\maketitle

\begin{abstract}
The scope of this paper is to compare two different approaches for solving the Wheeler-DeWitt (WDW) equation in the presence of homogeneous matter (inflaton) and perturbations around it. The standard Born-Oppenheimer (BO) decomposition, which consists of factorizing out the gravitational wave function and then defining the flow of the time through it, and a more general BO decomposition where the whole minisuperspace wave function is factorized out. The two approaches are compared, for simplicity, in the case of a minimally coupled inflaton with a flat potential. The consistency of the latter decomposition is checked against the former by comparing the resulting perturbation (Mukhanov-Sasaki) equations. Finally a few solutions to the homogeneous WDW not suitable for the traditional BO treatment are presented and the corresponding Mukhanov-Sasaki equations are evaluated. 
\end{abstract}

\section{Introduction}
The quest for a theory of quantum gravity is one of the biggest challenges of modern theoretical physics. While classical General Relativity (GR) has been able to successfully explain many astrophysical phenomena in the last century, it now appears to be ready for a further development. A full quantum mechanical description of the gravitational interaction is necessary for a complete understanding of a plethora of recent observations ranging from the features of the remnants of cosmological inflation to the evidence for dark matter and dark energy and the new discoveries on black holes physics unveiled by gravitational waves observations.\\
A consolidated, fundamental description of gravity at Planckian energies is certainly missing and all the diverse alternatives, from string theory to corpuscular models, suffer from mathematical inconsistencies, lack of predictability, or are simply not suitable for explaining all the physics we observe.
Canonical quantum gravity \cite{DeWitt} is one of the earliest theoretical attempts to combine GR with quantum mechanics. It consists of expressing the classical dynamics of the matter-gravity system in the canonical formalism and quantising it. The diffeomorphism invariance of classical GR leads to a series of constraints at the quantum level for the universe wave function: the Wheeler-DeWitt (WDW) equation, associated with time re-parametrisation invariance, and three momentum constraints related to the invariance with respect to spatial coordinates parametrisation. When the formalism is applied to homogenous matter-gravity systems, such as those relevant for cosmology, the momentum constraints are absent and one is simply left with the WDW equation. As one would expect from the quantisation of a theory, such as GR, which is invariant with respect to the choice of space-time coordinates, the WDW equation depends only on gravitation and matter degrees of freedom. Its application in a cosmological context is generally obtained by simply considering the minisuperspace coordinates as degrees of freedom and leads to a partial differential equation having the form of a time independent Schr\"odinger equation.\\
The WDW equation is a quantum constraint which arises from the canonical quantisation of GR as a gauge field theory. Therefore, while it is commonly believed that well above Planckian energy, the geometric description of space and time should break down and must be replaced by some new theory, one still expects that the well established formalism of quantum field theory applied to GR leads to a consistent theoretical description of the gravitational interactions almost up to Planck scales. It is then suitable for describing an early universe phenomenom such as inflation \cite{inflation}.
Despite QFT being a consolidated framework its application to cosmology suffers from conceptual problems. One of the most debated is the problem of time: in ordinary quantum mechanics, time is a ``classical'' parameter and the ``clock'' which measures time is therefore external to the quantum system. Moreover a quantum system evolves non-trivially in time when its different energy eigenstates are superimposed. The WDW equation is a constraint for the total energy of the universe which is set to zero. Thus it is not only impossibile conceptually to think of a ``clock'' external to the universe, but even if one assumed the existence of such a clock then a trivial time evolution for the system would be observed.\\
On a more abstract level the problem of the interpretation of the wave function of the universe is also posed. The Copenhagen interpretation of quantum mechanics assumes that the operation of measurement of some property of a system makes its wave function collapse in a non-deterministic way. The probabilistic interpretation of quantum mechanics has been proven to be successful when it is applied to a large number of quantum systems. Such an interpretation does not necessarily hold for the wave function of the universe, i.e. for a quantum mechanical system which is not microscopic, with no measurement apparatus external to it, and only one realisation of the universe is to be tested. Moreover the universe wave function solution to the WDW equation is not always normalisable.
It is often claimed that the WDW equation is an elegant quantum equation for the universe wave function which leads to no really quantitative predictions. Nonetheless the universe wave function contains information on how the universe began and how it expands evolving from a state with quantum properties to a state which has an increasingly classical behaviour.\\
When the inhomogeneities \cite{BOpert} are accounted for, the mathematical complexity of the WDW equation increases but they alleviate some of the interpretational issues listed so far: we live in a inhomogeneous universe and its inhomogeneities are what we actually observe \cite{pert,Planck}. Even the observables associated with the homogeneous universe, such as the Hubble parameter, are measured indirectly through the observation of the perturbations (elementary particles, photons, gravitational waves). To this extent the inhomogeneities also encode much information on the homogeneous (quantum) state of the Universe.\\
In order to introduce time evolution in the matter-gravity system a method originally applied to atoms and molecules \cite{BO} has been proposed \cite{BOcosm}. It is based on the Born-Oppenheimer (BO) decomposition and consists in factorising the full universe wave function into a ``heavy'' (or ``slow'') part, usually associated with the gravitational scale factor and a ``light'' (or ``fast'') part consisting of the matter degrees of freedom which is further factorizable. One is then led to a system of coupled equations: the gravity equation which determines the wave function of the gravitational scale factor and the matter equations for the homogeneous matter and its perturbations. In particular the equations for the perturbations take the form of the Mukhanov-Sasaki equations plus the quantum gravitational effects. Formally, for each matter degree of freedom, one finds a matter equation which formally resembles a ``time'' dependent Schr\"odinger equation. In contrast, the gravity equation depends on the back-reaction of matter on the gravitational degree of freedom and is the quantum counterpart of the semiclassical Friedmann equation. The equations contain contributions which, in the standard BO approach, are associated with non-adiabatic effects and, in this context, represent part of the quantum gravitational effects. Time is introduced in the matter equations by the probability flux associated with the gravitational wave function, and can be defined for gravity in the quasi-classical state or in a quantum state with definite momentum (plane-wave). In such an introduction more non-negligible quantum gravitational effects may arise.\\
The BO factorisation of the gravitational wave function is not always possibile since solutions may exist which are not factorizable or because the homogeneous matter wave function is not normalisable \cite{IGwdw}. Moreover factorising the matter and gravity homogeneous degrees of freedom in the same ``minisuperspace'' wave function allows for more control of the non-adiabatic effects necessarily originating from the decomposition. However, in this case, the emergence of time in the equation for the perturbations, should be treated more carefully as ambiguities in the procedure arise. The effects of the minisuperspace wave function on evolution of the perturbations has been already discussed previously. In \cite{kiefer} an attempt to evaluate the quantum gravitational corrections originating from the minisuperspace wave function was illustrated. In a more recent paper \cite{bounce} the issue was discussed for slow-roll inflation and in the large limit for the scale factor. With such assumption the leading order quantum corrections were estimated. Finally, in \cite{newBO}, a procedure to estimate the quantum gravitational corrections was illustrated for a non-minimally coupled inflaton field and a de Sitter expansion.\\
In this paper we shall discuss and compare how time can be introduced in the matter-gravity system for a few cases which can be treated almost exactly, i.e. with a minimally coupled inflaton and a constant potential. These cases are very interesting as they can be solved exactly at the classical level as well and the corresponding exact solutions for the homogeneous WDW equations can also be found. Furthermore the quantum gravitational corrections to the Mukhanov-Sasaki equation can be calculated both using the traditional BO decomposition, i.e. factorising out the gravity wave function, and a more general BO decomposition which consists in factorising the minisuperspace degrees of freedom. The consistency of the second approach can be tested against the results obtained from the traditional one.\\
The paper is organised as follows. In Section 2 the general formalism is illustrated and the relevant equations obtained and discussed. In Section 3 we find, through diverse approaches, different solutions to the homogeneous WDW equation. These solutions are the used, in Section 4, to introduce a time for the MS equation. Furthermore the quantum gravitational corrections along the classical trajectory for such an equation are evaluated. The traditional BO decomposition is then compared with the more general one and the resulting equations are compared and discussed.

\section{The model}
Let us consider a minimally coupled scalar field (inflatonic matter) on a curved, spatially flat, spacetime described by the following action
\be{act0}
S=\int d^4 x\sqrt{-g}\paq{-\frac{\M^2}{2}R+\frac{1}{2}\partial_\mu\varphi\partial^{\mu}\varphi-V(\varphi)}
\ee
where $\M$ is the Planck mass, $R$ is the Ricci scalar and $\varphi$ is the inflaton field.
The action can be decomposed into a homogeneous part plus fluctuations around it. Let $a$ and $\phi$ be the scale factor and the homogeneous part of the inflaton field $\varphi$. The fluctuations will be described in terms of a single, Mukhanov-Sasaki (MS) field $v(x,t)$.
The full Lagrangian density is then given by
\be{act} 
\mathcal{L}=-L^3\pa{3\M^2\frac{a\dot a^2}{N}-\frac{a^3\dot \phi^2}{2N}+a^3NV}+\sum_{k\neq 0}\mathcal{L}_k
\ee
where the dot denotes the derivative w.r.t. a generic time variable, $N$ is the corresponding lapse function and $\mathcal{L}_k$ is the Lagrangian of each k-mode of the MS field. Let us note that, on a flat 3-space and considering both homogeneous and inhomogeneous quantities, one must introduce an unspecified length $L$ (see \cite{BOpert} for more details) and we shall set to $L=1$ in what follows. When the conformal time ($N=a$) is chosen, the Lagrangian of the perturbations takes the form 
\be{pertslag}
\mathcal{L}_k=\frac{1}{2}\pa{v_k'^2-\omega_k^2v_k^2}
\ee
with
\be{freqdef}
\omega_k^2=k^2-\frac{z''}{z}.
\ee
The time dependent mass terms for the scalar and tensor perturbations, $z''/z$, are defined in terms of the homogeneous, classical degrees of freedom by
\be{zdefsc}
z\equiv a\sqrt{\ep{1}},\; {\rm with}\; \ep{1}\equiv -\frac{H'}{a H^2}\; {\rm and }\; H=\frac{a'}{a^2}
\ee
for the scalar case and by
\be{zdeftn}
z\equiv a
\ee
for the tensor case. Moreover the MS variable $v_k$ describing the scalar sector is given by
\be{vk}
v_k=\frac{z\,a'}{a\,\phi'}\delta\phi_k
\ee
in the uniform curvature gauge, where $\delta\phi_k$ is the Fourier transform of the inflaton field fluctuations. For the tensor sector 
\be{vkt}
v_k=\frac{a\M}{\sqrt{2}}h_k
\ee
where $h_k$ is the Fourier transform of each polarisation of the gauge invariant perturbations of the homogeneous metric. Let us note that the scalar and the tensor perturbations are decoupled at the linear order and therefore can be studied separately. 
\\
The system Hamiltonian is constrained as a consequence of GR time re-parametrisation invariance, i.e.
\be{ham}
{\mathcal H}=\frac{\pi_\phi^2}{2a^2}-\frac{\pi_a^2}{2\m^2}+a^4 V+\sum_{k\neq0} \mathcal H_k=0
\ee
where
\be{Hk}
\mathcal{H}_k=\frac{1}{2}\left(\pi_k^2+\omega_k^2v_k^2\right)
\ee
is the Hamiltonian for the mode $k$ of each perturbation $v_k$. Let us note that the classical momenta are
\be{momenta}
\pi_a=-6\M^2 a',\;\pi_\phi=a^2\phi',\;\pi_k=v'_k.
\ee
Once quantised the Hamiltonian constraint leads to the Wheeler-DeWitt equation.
On quantising the Hamiltonian constraint and allowing for different orderings of the kinetic term for gravity, the resulting WDW equation takes the form
\be{wdw0}
\paq{\frac{1}{2\m^2}a^{-l}\partial_a a^{-j}\partial_a a^{-i}-\frac{1}{2a^2}\partial_\phi^2+a^4 V+\sum_k \hat{\mathcal H}_k}\Psi(a,\phi,\pag{v_k})=0
\ee
with $i=-l-j$. Henceforth we shall discuss two diverse (and commonly adopted) orderings: the case $i=l=j=0$ and the case $i=0$ and $l=-j=1$. Finally the results for the latter case will be presented and discussed in detail.
\subsection{Born-Oppenheimer decomposition}
The traditional BO approach to the inflaton-gravity system consists in factorising the total wave function into a gravitational wave function $\psi(a)$ and a matter wave function $\chi_M(a,\phi,\pag{v_k})$ which can be further decomposed as follows
\be{tradBO}
\Psi\pa{a,\phi,\pag{v_k}}=\psi(a)\chi_M(a,\phi,\pag{v_k})=\psi(a)\chi_0(a,\phi)\prod_{k\neq 0}\chi_k(a,v_k).
\ee
On projecting out the matter wave function, one is then led to a gravity equation
\be{greq}
\partial_a^2\tilde \psi-\frac{j}{a}\partial_a\tilde \psi+2\m^2\av{\hat {\mathcal H}_0}\tilde \psi
=\av{\partial_a\tilde \chi_0|\partial_a \tilde \chi_0}\tilde \psi,
\ee
where
\be{defO}
\av{\hat O}=\int_{-\infty}^{+\infty}\pa{\prod_{k\neq 0}\rd v_k}\rd \phi\,\chi_M^*\hat O\chi_M,
\ee
and $\tilde \chi_k$ is a properly rephased wave function for each k-mode (see below for the exact definition) with $k=0$ indicating that of the homogeneous scalar field. For simplicity, we neglected the contributions originating from the perturbations $v_k$ and $\hat {\mathcal H}_0$ is the Hamiltonian of the homogeneous scalar field. Furthermore the projection of the WDW along $\prod_{p\neq k}\langle \chi_{p}|$ leads to a set of equations for the matter part of the system of the following form
\be{mateq}
\paq{2\frac{\partial_a\tilde \psi}{\tilde \psi}-\frac{j}{a}}\partial_a\tilde \chi_k+2\m^2\pa{\hat {\mathcal H}_k-\av{\hat{\mathcal H}_k}}\tilde \chi_k+\pa{\partial_a^2-\av{\tilde \partial_a^2}}\tilde \chi_k=0
\ee
where 
\be{deftildeO}
\av{\hat {\tilde O}}=\int_{-\infty}^{+\infty}\pa{\prod_{k\neq 0}\rd v_k}\rd \tilde\phi\,\chi_M^*\hat O\tilde \chi_M,
\ee
and $\hat{\mathcal H}_k$ is the Hamiltonian for a mode. Eq. (\ref{mateq}) is an exact partial differential equation involving only the scale factor $a$ and one matter degree of freedom (either $\phi$ or $v_k$). By construction eq. (\ref{mateq}) implies $\langle\tilde\chi_k |\partial_a|\tilde \chi_k\rangle=0$. It consists of a contribution which depends on the gravity wave function and is associated with the introduction of time, the Hamiltonian for the corresponding wave function and second derivatives with respect to the scale factor. The latter contributions are usually called ``non-adiabatic'' effects and, in this context, they generate some of the quantum gravitational corrections to the semiclassical evolution of the MS field. The remaining quantum gravitational corrections emerge from the gravity wave function and the introduction of time.\\ 
Instead of the approximate eq. (\ref{greq}), an exact gravitational equation containing the contributions of each perturbation $v_k$ is generally obtained by the BO decomposition. At the semiclassical level such contributions are also expected and are responsible for the back-reaction of the perturbations on the homogeneous scale factor. When the energy density of the homogeneous part dominates, such as during inflation, the back-reaction effects can be safely neglected. We adopt here this, physically reasonable, assumption. \\
Let us note that the ``tilded'' wave functions for matter and gravity which appear in the above equations are related to those introduced in the BO factorisation (\ref{tradBO}) by the following phase redefinition:
\be{phred}
\tilde \psi=\exp\paq{\int da'\av{\partial_a}}\psi,\quad \tilde \chi_M=\exp\paq{-\int da'\av{\partial_a}}\chi_M
\ee
and
\be{phredi}
\tilde \chi_k=\exp\paq{-\int da'\av{\partial_a}_k}\chi_k,
\ee
where 
\be{dak}
\av{\partial_a}_k=\int_{-\infty}^{+\infty}\rd v_k\chi_k^*\partial_a\chi_k
\ee
and
\be{avda}
\av{\partial_a}=\sum_{k=0}^{\infty}\av{\partial_a}_k\simeq \av{\partial_a}_0\equiv \langle \chi_0|\partial_a|\chi_0\rangle.
\ee
The last approximation consists in, again, neglecting the contributions of the perturbations with respect to the homogeneous part. Such an approximation can be justified just as the neglection of the back-reaction of the inhomogeneities in the gravity equation. Therefore the product 
\be{prodhom}
\tilde\Psi_0\equiv \tilde \psi \,\tilde \chi_0=\psi\,\chi_0,
\ee
satisfies the WDW equation in the minisuperspace (homogeneous) approximation, as can be easily checked simply by summing (\ref{greq}) and (\ref{mateq}) with $k=0$. Let us finally note that the projections on the matter wave functions employed to obtain (\ref{greq}) and (\ref{mateq}) from the WDW equation are mathematically acceptable if each wave function in the factorisation of the matter part is normalizable.\\
A different procedure for decomposing \`a la Born-Oppenheimer the total wave function $\Psi(a,\phi,\{v_k\})$ was recently adopted and discussed for the WDW equation of an inflaton-gravity system in the presence of a non-minimal coupling (and in particular for the induced gravity case) \cite{newBO}. The following factorisation was proposed: 
\be{BOdec}
\Psi(a,\phi,\pag{v_k})=\tilde \Psi_0(a,\phi)\prod_{k\neq 0}\tilde\chi_k(a,\phi,v_k).
\ee
A procedure analogous to that illustrated above for the traditional BO approach, can be applied to eq. (\ref{wdw0}). Projecting out the inhomogeneous matter wave function and neglecting the back-reaction of the perturbations on the homogeneous modes, leads to the following homogeneous WDW equation for the minisuperspace degrees of freedom
\be{hWDW}
\paq{G^{\alpha\beta}\partial_\alpha\partial_\beta-\frac{1+j}{2}\partial_A+\m^{-4}\re^{6A} V(F)}\tilde \Psi_0(A,F)=0
\ee
where we redefined the homogeneous degrees of freedom as $A=\ln \pa{\m a}$, $F=\phi/\m$, and introduced the super-metric,
\be{Guu}
G\equiv\frac{1}{2}
\begin{pmatrix}
1 & 0\\ 0&-1
\end{pmatrix},
\ee
with the Greek indices $(\alpha,\beta)$ having values $\pag{A,F}$.
Each perturbation then obeys the following equation:
\be{keq}
\pag{G^{\alpha\beta}\paq{2\frac{\partial_\alpha\tilde \Psi_0}{\tilde \Psi_0}\partial_\beta+\pa{\partial_\alpha\partial_\beta-\av{\partial_\alpha\partial_\beta}}}+\frac{1+j}{2}\partial_A+\re^{2A}\pa{\hat{\mathcal H}_k-\av{\hat{\mathcal H}_k}}}\tilde \chi_k=0.
\ee
Such a procedure was proposed in order to overcome the problem of having non-factorizable solutions of the homogeneous WDW. The presence of both the minisuperspace degrees of freedom in the perturbation equation has important consequences on the definition of the quantum gravitational corrections associated with the non-adiabatic terms. In the following sections we illustrate a procedure to define such corrections consistently with that employed within the traditional BO factorisation (\ref{tradBO}). 
\section{Constant potential case}
Henceforth we shall restrict our analyses to the constant potential case, i.e. $V=\lambda \m^4$. 
Moreover we make the $j=-1$ ordering choice. Consideration of such a simplified case is sufficient to illustrate a method whereby the two different BO decompositions may be compared with each other.\\
The WDW equation in the minsuperspace (\ref{hWDW}) then takes the form:
\be{hWDW2}
\paq{G^{\alpha\beta}\partial_\alpha\partial_\beta+\lambda \re^{6A}}\tilde \Psi_0(A,F)=0
\ee
and can be solved exactly. With the ansatz
\be{ansfacds}
\tilde \Psi_0(A,F)=\re^{iPF}\tilde\psi(a)
\ee
Eq. (\ref{hWDW2}) takes the form
\be{hWDW3}
\pa{\partial_A^2+P^2+2\lambda \re^{6A}}\tilde\psi(a)=0
\ee
and has the following general solution:
\be{solfa}
\tilde\psi(a)=\pag{c_1 H_{\bar\nu}^{(1)}\pa{B\, \m^3 a^3}+c_2 H_{\bar\nu}^{(2)}\pa{B\, \m^3 a^3}},
\ee
where 
\be{defsHankel}
\bar \nu=-\frac{iP}{3},\; B=\sqrt{\frac{2\lambda}{9}},
\ee
and $H_{\nu}^{(1)}$ and $H_{\nu}^{(2)}$ are the Hankel functions. 
Let us note that $A$ and $F$ are factorized in the solution (\ref{ansfacds}) and we therefore expect that $\tilde \chi_0\equiv \re^{iPF}$ and $\tilde \psi(a)$ satisfy equations (\ref{greq}) and (\ref{mateq}) as can be easily verified.
\subsection{Non-factorizable solutions}
A different technique may be adopted to find the solutions of (\ref{hWDW2}). Let us consider the following change of variables $(A,F)\rightarrow (u,v)$ with
\be{chvaruv}
u=\re^{3\pa{A+F}},\quad v=\re^{3\pa{A-F}},
\ee
then Eq. (\ref{hWDW2}) then takes the form
\be{hWDW2c}
\paq{18\partial_u\partial_v+\lambda}\tilde\Psi_0=0
\ee
and its solution is
\be{solnew}
\tilde \Psi_0=\exp\paq{\pm i\pa{A\, u+B\,v}}
\ee
with
\be{ABcond}
A\,B=\frac{\lambda}{18}.
\ee
In terms of $a$ and $\phi$ the solution becomes
\be{solnewAF2alzero}
\tilde \Psi_0=\exp\paq{\pm i\pa{\m a}^3\sqrt{\frac{2\lambda}{9}}\cosh\pa{3\frac{\phi}{\m}+\frac{1}{2}\ln\frac{\lambda}{18 B^2}}}
\ee
where $B$ is a free parameter and we choose it to be a real number.
Let us note that (\ref{solnewAF2alzero}) cannot be factorized into a part depending on $a$ and a part function of $a$ and $\phi$. Despite this latter solution not being naturally factorizable  \'a la BO, one may still look for a gravity wave function $\tilde \psi(a)$ satisfying (\ref{greq}) and a homogeneous matter wave function $\tilde \chi(a,\phi)$ , which satisfies (\ref{mateq}), defined as
\be{defpsi}
\tilde \Psi_0= \tilde \psi\,\pa{\tilde \psi^{-1}\tilde \Psi_0}\equiv \tilde \psi\,\tilde \chi.
\ee
If the decomposition (\ref{defpsi}) were possible then ``time'' could be introduced by exploiting the gravitational wave function $\tilde \psi(a)$ only. Moreover, in such a case, $\tilde \chi_0$ should also satisfy the identity $\av{\tilde \partial_a}_0=0$ i.e.
\be{idenver}
0\stackrel{?}{=}\av{\tilde \partial_a}_0=\frac{1}{\tilde \psi}\int_{-\infty}^{+\infty}\rd\phi\paq{\frac{3i}{a}\pa{\m a}^3\cosh\pa{3\frac{\phi}{\m}+\frac{1}{2}\ln\frac{\lambda}{18B^2}}-\frac{\partial_a\tilde \psi}{\tilde \psi}}.
\ee
One can easily see that the above request cannot be satisfied by any choice of $\tilde \psi$ and therefore one must conclude that a different procedure for the introduction of time in the equation for the perturbation, based on the full set of minisuperspace variables, is needed. \\
Let us note that the homogeneous WDW equation cast in the form (\ref{hWDW2c}) reveals a few symmetries not obvious in its original form (\ref{hWDW2}). The symmetry operators
\be{symm}
\hat X_1=\partial_u,\;\hat X_2=\partial_v,\; \hat X_3=u\partial_u-v\partial_v
\ee
can generate new solutions starting from those already found. In particular we observe that $\hat X_{1,2}$ generate new, non-trivial, solutions when acting on (\ref{ansfacds}) while $\hat X_3$ leaves the set of solutions (\ref{ansfacds}) invariant and, in contrast, $\hat X_3$ generates new solutions when acting on (\ref{solnew}), which is indeed invariant with respect to the symmetry transformations $\hat X_{1,2}$.\\
In terms of $A$ and $F$ the above symmetry operators are given by
\be{symmAF}
\hat X_1=\frac{\re^{-3(A-F)}}{6}\pa{\partial_A-\partial_F},\;\hat X_1=\frac{\re^{-3(A+F)}}{6}\pa{\partial_A+\partial_F},\;\hat X_3=\frac{1}{3}\partial_F.
\ee
For example, starting from (\ref{ansfacds}) with $c_1=1$ and $c_2=0$ one finds
\be{x1}
\hat X_1 \Psi_0=\frac{\sqrt{2\lambda}}{2}\re^{iPF+3F}\pa{H_{-1+\nu}^{(1)}\pa{B\re^{3A}}-H_{1+\nu}^{(1)}\pa{B\re^{3A}}}-3\nu \re^{-3\pa{A-F}}\Psi_0
\ee
where the new solution (\ref{x1}) is no longer factorizable into a gravity and a matter part and is not normalisable with respect to the homogeneous inflaton field. Similarly $\hat X_3$ applied to the solution (\ref{solnewAF2alzero}) can generate the new solution
\be{x3}
\hat X_3 \Psi_0=\pm i\sqrt{2\lambda}\pa{\m a}^3\sinh\pa{3\frac{\phi}{\m}+\frac{1}{2}\ln\frac{\lambda}{18 B^2}}\Psi_0
\ee
which is neither factorizable nor normalizable.
Let us note that application of the operators (\ref{symm}) to exact solutions may, in general, generate new, independent, solutions or reproduce some of the solutions already found. We shall not discuss the general features of this approach in this paper but limit ourselves, in the next section, to some discussion as to how these different solutions influence the introduction of time and the quantum gravitational correction associated with it.

\subsection{Quantum gravitational corrections}
If one now follows the traditional approach based on the decomposition (\ref{tradBO}) the matter equation can be cast on the following form:
\be{mateqA}
\paq{\re^{-2A}\frac{\partial_A\tilde\psi}{\tilde \psi}\partial_A+\pa{\hat {\mathcal H}_k-\av{\hat {\mathcal H}_k}}+\frac{\re^{-2A}}{2}\pa{\partial_A^2-\av{\tilde \partial_A^2}}}\tilde \chi_k=0
\ee
and ``time'' can be introduced by identifying 
\be{timetrad}
\re^{-2A}\frac{\partial_A\tilde\psi}{\tilde \psi}\partial_A=-i\partial_\eta+ \Delta q\,\partial_A, 
\ee
where $\Delta q\,\partial_A$ is the operator which contains the quantum gravitational corrections emerging from the definition of time. When such corrections are negligible, the ``classical'' relation between $A$ and the conformal time $\eta$ should be reproduced, i.e.
\be{etaArel}
\partial_\eta=(\partial_\eta A)_{\rm cl}\partial_A=\pa{a H}_{\rm cl}\partial_A\equiv -\re^{-2A}\pi_{A,{\rm cl}}\partial_A,
\ee
where $\pi_{A,{\rm cl}}$ is the classical momentum associated with $A$. Thus, the following relation must hold
\be{cltimetrad}
i\re^{-2A}\frac{\partial_A\tilde\psi}{\tilde \psi}=\frac{i}{\m^2a}\frac{\partial_a\tilde\psi}{\tilde \psi}=\pa{aH}_{\rm cl},
\ee
where the subscript ``cl'' indicates that the quantities are evaluated on the classical trajectory.
Let us note that, while the definition (\ref{timetrad}) can always be written, the introduction of time is physically acceptable only if the quantum corrections are small and can be treated as a perturbation. If this were not the case the matter equation would remain timeless and its solution would be associated with a fully quantum regime. \\
If we consider the exact solution (\ref{solfa}) with $c_2=0$ and take the large $a$ limit then
\be{cltimetrad2}
\frac{i}{\m^2a}\frac{\partial_a\tilde\psi}{\tilde \psi}=\sqrt{2\lambda}\,a\m-\frac{3i}{2a^2\m^2}+\frac{9+4 P^2}{8\sqrt{2\lambda}a^5\m^5}+\mathcal{O}\pa{\frac{1}{\m^6 a^6}}.
\ee
From the classical Friedmann equation, and in the same ($\m a\gg 1$) limit one finds
\be{ahlim}
(a H)_{\rm cl}=\sqrt{2\lambda}\, a\m+\frac{P^2}{2\sqrt{2\lambda}\,a^5\m^5}+\mathcal{O}\pa{\frac{1}{\m^6 a^6}},
\ee
which, on comparing with (\ref{cltimetrad2}), gives
\be{dQtrad}
\Delta q=-\frac{3}{2a^2\m^2}+\mathcal{O}\pa{a^{-5}},
\ee
and therefore one has
\be{tradtime}
\re^{-2A}\frac{\partial_A\tilde\psi}{\tilde \psi}\partial_A\simeq-i\partial_\eta-\frac{3i}{2\m^2\pa{a^3H}_{\rm cl}}\,\partial_\eta.
\ee
One can now express the quantum gravitational corrections associated with the non-adiabatic effects in terms of $\partial_\eta$: given (\ref{etaArel}) one easily obtains
\be{nonadiabtrad}
\partial_A^2=\frac{1}{\pa{a H}^2}\paq{\partial_\eta^2-aH\pa{1-\ep{1}}\partial_\eta},
\ee
where the subscript indicating the classical trajectory has been removed and $\ep{1}\equiv -H'/\pa{aH^2}$ is the slow-roll parameter associated with the variation of the Hubble parameter. Once ``time'' is introduced the matter equation (\ref{mateq}) takes the following final form
\be{mateqtrad}
\pag{-i\partial_\eta+\pa{\hat {\mathcal H}_k-\av{\hat {\mathcal H}_k}}+\hat Q-\av{\hat {Q}}}\tilde \chi_k=0,
\ee
where
\be{defQ}
\hat Q=\frac{1}{2\m^2a^4H^2}\paq{\partial_\eta^2-aH\pa{4-\ep{1}}\partial_\eta}
\ee
describes the quantum gravitational corrections. On neglecting such corrections and properly re-phasing the matter wave function $\tilde \chi_k$
\be{reph}
\tilde \chi_k\rightarrow \tilde \chi_{k,s}=\exp\pa{-i\int^\eta \rd \eta' \av{\hat {\mathcal H}_k}}\tilde \chi_k
\ee
one is finally led to the standard Mukhanov-Sasaki equation (in the Schr\"odinger representation):
\be{tradMS}
\pa{-i\partial_\eta+\hat {\mathcal H}_k}\tilde \chi_k=0.
\ee
In contrast, if one follows the approach based on the BO factorisation (\ref{BOdec}), the matter equation has the form
\be{keq2}
\pag{\re^{-2A}G^{\alpha\beta}\paq{2\frac{\partial_\alpha\tilde \Psi_0}{\tilde \Psi_0}\partial_\beta+\pa{\partial_\alpha\partial_\beta-\av{\partial_\alpha\partial_\beta}}}+\pa{\hat{\mathcal H}_k-\av{\hat{\mathcal H}_k}}}\tilde \chi_k=0
\ee
and ``time'' can be introduced through the following definition
\be{timenew}
2\re^{-2A}G^{\alpha\beta}\frac{\partial_\alpha\tilde \Psi_0}{\tilde \Psi_0}\partial_\beta=-i\partial_\eta+\pa{\Delta q}^{\alpha}\partial_\alpha.
\ee
The quantum corrections are now given by $\pa{\Delta q}^{\alpha}\partial_\alpha$ and the definition of ``time'' along the classical trajectory in minisuperspace is given by 
\be{etaAFrel}
\partial_\eta=\pa{\partial_\eta A}\partial_A+\pa{\partial_\eta F}\partial_F=-2\re^{-2A}G^{\alpha\beta}\pi_{\alpha,{\rm cl}}\partial_\beta,
\ee
where $\pi_{\alpha,{\rm cl}}$ are the classical momenta associated with the homogeneous degrees of freedom $A$ and $F$.\\
If we now consider the exact (factorizable) solution (\ref{ansfacds}) then ``time'' can be easily introduced as follows:
\be{timnew2}
2\re^{-2A}G^{\alpha\beta}\frac{\partial_\alpha\tilde \Psi_0}{\tilde \Psi_0}\partial_\beta=-i (aH) \partial_A+\Delta q \,\partial_A-i\frac{P}{\m a^2}\partial_F=-i\partial_\eta+\Delta q \,\partial_A,
\ee
where $P/(\m a^2)=F'_{\rm cl}$ and $\Delta q$ is simply given by (\ref{dQtrad}). Therefore, in both the BO approaches, one finds exactly the same form for the quantum gravitational corrections which emerge from the introduction of time. One must finally evaluate the non-adiabatic corrections i.e.
\be{nonadiabnew}
\re^{-2A}G^{\alpha\beta}\pa{\partial_\alpha\partial_\beta-\av{\partial_\alpha\partial_\beta}}\tilde \chi_k=\frac{1}{2\m^2a^2}\pa{\partial_A^2-\av{\partial_A^2}-\partial_F^2+\av{\partial_F^2}}\tilde \chi_k
\ee
and, in particular, these latter corrections with the assumption that $\tilde \chi_k$ is a function of $\eta$. Such a request is not the only possibility offered by this approach where the non-adiabatic corrections are associated with two independent vector fields ($\partial_A$ and $\partial_F$). However, this assumption is necessary in order to compare the resulting corrections with those obtained previously with the traditional BO approach.\\
From the definition (\ref{etaAFrel}) and given that
\be{clsol}
\pi_{A,{\rm cl}}=-\sqrt{2\lambda \re^{6A}+P^2},\quad \pi_{F,{\rm cl}}=P,
\ee
one has
\be{timeAF}
\partial_\eta=\re^{-2A}\pa{\sqrt{2\lambda \re^{6A}+P^2}\partial_A+P\partial_F},
\ee
which defines the congruence of trajectories defining the time flow in each point of the minisuperspace. This congruence defines two vector fields:
\be{dTtaudef}
\left\{
\begin{array}{l}
\partial_T=\re^{-6A}\pa{\sqrt{2\lambda \re^{6A}+P^2}\partial_A+P\partial_F}\\
\partial_\tau=\re^{-6A}\sqrt{2\lambda \re^{6A}+P^2}\pa{P\partial_A+\sqrt{2\lambda \re^{6A}+P^2}\partial_F}
\end{array}\right.
\ee
orthogonal to each other (with respect to the super-metric $G^{\alpha\beta}$) and commuting $\paq{\partial_T,\partial_\tau}=0$, i.e. with $\partial_T \tau=0$. The global reparametrisation of minisuperspace in terms of $\pa{T,\tau}$ is obtained by the following change of variables:
\be{chvar}
\left\{\begin{array}{l}
A=\frac{1}{6}\ln\paq{\frac{36\lambda^2\pa{T+P\tau}^2-P^2}{2\lambda}}\\
F=2\lambda \tau-\frac{1}{3}\arctanh\paq{\frac{6\lambda\pa{T+P\tau}}{P}}
\end{array}\right.
\ee
The vector field $\partial_\eta$ is, by construction, parallel to $\partial_T$
\be{detaTt}
\partial_\eta=\pa{\partial_\eta T}\partial_T+\pa{\partial_\eta \tau}\partial_\tau=\re^{4A} \partial_T
\ee
and therefore 
\be{eqeta}
\partial_\eta T=\re^{4A}=\paq{\frac{36\lambda^2\pa{T+P\tau}^2-P^2}{2\lambda}}^{2/3},\;\partial_\eta \tau=0
\ee
i.e. $\tau$ is constant on varying $\eta$. Thus, the first relation in (\ref{eqeta}) can be solved obtaining 
\be{etaTtau}
\eta-\eta_0=-\frac{2^{2/3}\lambda(T+P\tau)}{P^2}\!\!\!\phantom{F}_2F_1\pa{\frac{5}{6},1;\frac{3}{2};\frac{36\lambda^2\pa{T+P\tau}^2}{P^2}}.
\ee
Let us note that this last expression defines $\eta$ on the minisuperspace.\\ It is now possible to calculate the derivatives with respect to $T$ and $\tau$ of any function of $\eta$:
\be{dertaufeta}
\partial_\tau f(\eta)=\pa{\partial_\tau\eta}\partial_\eta f(\eta)=P\partial_T f(\eta)
\ee
and
\be{derTfeta}
\partial_T f(\eta)=\pa{\partial_T\eta}\partial_\eta f(\eta)=\paq{\frac{2\lambda}{36\lambda^2(T+P\tau)^2-P^2}}^{2/3}\partial_\eta f(\eta),
\ee
where, from (\ref{chvar}), one has
\be{auxrel}
\re^{-6A}=\frac{2\lambda}{36\lambda^2(T+P\tau)^2-P^2}.
\ee
On inverting the relations (\ref{dTtaudef}) 
\be{dAFdef}
\left\{
\begin{array}{l}
\partial_A=\frac{\sqrt{2\lambda \re^{6A}+P^2}}{2\lambda}\pa{\partial_T-\frac{P}{2\lambda\re^{6A}+P^2}\partial_\tau}\\
\partial_F=-\frac{1}{2\lambda}\pa{P\partial_T-\partial_\tau}
\end{array}\right.
\ee
and using the results (\ref{dertaufeta}), (\ref{derTfeta}), (\ref{auxrel}) one finds
\be{dAFfeta}
\left\{
\begin{array}{l}
\partial_A f(\eta)=\frac{\re^{2A}}{\sqrt{2\lambda \re^{6A}+P^2}}\partial_\eta f(\eta)\\
\partial_Ff(\eta)=0
\end{array}\right..
\ee
In terms of these derivatives one can finally calculate the non-adiabatic corrections to the matter equation (\ref{nonadiabnew}) along the classical trajectory. Given the classical relations
\be{clrels}
H^2=\m^2\re^{-6A}\pa{P^2+2\lambda \re^{6A}},\quad \ep{1}=\frac{3P^2}{P^2+2\lambda \re^{6A}}
\ee
one finds
\be{nonadneweta}
\re^{-2A}G^{\alpha\beta}\partial_\alpha\partial_\beta \tilde \chi(\eta)=\frac{1}{2a^4\m^2H^2}\paq{\partial_\eta^2-aH\pa{1-\ep{1}}\partial_\eta} \tilde \chi(\eta)
\ee
and the matter equation (\ref{mateqtrad}) is recovered. Let us note that, the procedure illustrated above, for the introductions of time and the evaluation of the non-adiabatic effects, reproduces the same results as the traditional BO decomposition. Nonetheless the procedure is more general and can be applied to the solutions of the homogeneous WDW equation (\ref{hWDW2}) which cannot be factorized into a gravity part and a matter part. For such a case the non-adiabatic effects are the same and differences, if any, emerge from the introduction of time. \\
Let us now consider the application of the procedure illustrated above to the solutions (\ref{solnewAF2alzero}), (\ref{x1}) and (\ref{x3}) which cannot be factorized. For such cases the BO factorisation (\ref{BOdec}) is necessary.\\ 
For (\ref{solnewAF2alzero}) the introduction of time according to (\ref{timnew2}) gives exactly $\Delta q=0$ and 
\be{timnew3}
2\re^{-2A}G^{\alpha\beta}\frac{\partial_\alpha\tilde \Psi_0}{\tilde \Psi_0}\partial_\beta=-i\partial_\eta.
\ee
The resulting matter equation must be modified accordingly and the leading quantum gravitational corrections only consist of the non-adiabatic effects, i.e.
\be{newQ}
\hat Q\rightarrow \hat Q_*=\frac{1}{2\m^2a^4H^2}\paq{\partial_\eta^2-aH\pa{1-\ep{1}}\partial_\eta}.
\ee
Two more non-factorizable solutions, (\ref{x1}) and (\ref{x3}), have been presented in the previous section as an example of the application of the symmetry operators (\ref{symm}) to the solutions previously found. For such solutions new quantum gravitational corrections arise when time is introduced and in particular one has
\be{qgcx1}
\Delta q^A=-\frac{3}{2a^2\m^2},\; \Delta q^F=-\frac{3}{a^2\m^2}
\ee
for (\ref{x1}) and
\be{qgcx3}
\Delta q^A=\frac{3}{a^2\m^2},\; \Delta q^F=-\frac{3aH}{P}
\ee
for (\ref{x3}), and therefore the corresponding total corrections are given by
\be{newQ1}
\hat Q_1=\hat Q=\frac{1}{2\m^2a^4H^2}\paq{\partial_\eta^2-aH\pa{4-\ep{1}}\partial_\eta}
\ee
and
\be{newQ1}
\hat Q_2=\frac{1}{2\m^2a^4H^2}\paq{\partial_\eta^2+aH\pa{5+\ep{1}}\partial_\eta}.
\ee
\section{Conclusions}
We compared two different BO decompositions for the WDW equation and checked the consistency of one against the other. For simplicity, we restricted our calculations, to the case of a minimally coupled inflaton with a constant potential. For the homogeneous case this system can be solved exactly (both at the classical and the quantum levels) with different methods and some of the resulting quantum solutions are presented and discussed. Moreover the quantum system can be easily decomposed \`a la BO following the traditional approach, i.e. on factorizing out the gravity wave function and then introducing a time with it, or can be decomposed starting with a more general ansatz, i.e. factorizing the wave function of the minisuperspace. In both cases exact solutions can be found.\\
The existence of these solutions allows us to perform an exact comparison of the two decompositions. In particular, the latter approach can be checked against the former when the minisuperspace wave function admits a non trivial BO decomposition and a gravitational wave function can be factorized from it.\\
Nonetheless the definition of the quantum gravitational corrections associated with non adiabatic effects appears more ambiguous in the second approach where the form of such corrections depends on both of the minisuperspace (homogeneous) degrees of freedom. In particular, in the traditional approach, once time is introduced through the gravitational wave function, one only calculates the quantum gravitational correction to the MS equation along the classical trajectory. In contrast, in the second approach based on the factorisation of the minisuperspace wave function, one could, in principle, also include the corrections orthogonal to the classical trajectory.\\
A prescription must be given in order to obtain the same results when the two methods are applied and the quantum corrections are restricted to those along the classical flow of the time. Here we illustrated how the general method must be correctly applied in order to reproduce the results obtained with the traditional one. Finally such a general method is applied to a few cases which are not amenable for the traditional treatment, i.e. to non factorizable solutions of the homogeneous WDW equation, and the corresponding quantum gravitational corrections are calculated.

\section{Acknowledgements}
Alexander Y. Kamenshchik is supported in part by the Russian Foundation for Basic Research grant No. 20-02-00411.


\end{document}